\documentclass{aa}  

\usepackage{graphicx}
\usepackage{txfonts}
\usepackage{textcomp}

\def\wasp{WASP-76\,b}

\def\kms{km\,s$^{-1}$}

\def\kp{$K_{\rm P}$}
\def\h2o{H$_2$O}
\def\nh3{NH$_3$}

\begin{document} 

\title{Searching for the origin of the Ehrenreich effect in ultra-hot Jupiters}
\subtitle{Evidence for strong C/O gradients in the atmosphere of \wasp?}
\titlerunning{Evidence for strong C/O gradients in the atmosphere of \wasp?}
\author{
A.\,S\'anchez-L\'opez\inst{1},
R.\,Landman\inst{1},
P.\,Molli\`ere\inst{2},
N.\,Casasayas-Barris\inst{1},
A.\,Y.\,Kesseli\inst{1}, and
I.\,A.\,G.\,Snellen\inst{1}
}

\institute{Leiden Observatory, Leiden University, Postbus 9513, 2300 RA, Leiden, The Netherlands \and Max-Planck-Institut f\"ur Astronomie, K\"onigstuhl 17, 69117 Heidelberg, Germany \\
\email{alexsl@strw.leidenuniv.nl}
}

\authorrunning{A.\,S\'anchez-L\'opez et al.}
%

\abstract
{Extreme temperature contrasts between the day and nightside of ultra-hot Jupiters result in significantly asymmetric atmospheres, with a large expansion occurring over a small range of longitude around the terminator. Over the course of a transit, \wasp\ rotates by about 30\textdegree, changing the observable part of the atmosphere and invoking variations in the appearance of its constituents. Specifically, during the latter part of the transit, the planet’s trailing limb probes an increasing portion of its inflated dayside, which has a higher atmospheric detectability in transmission. As recently reported, this results in time-variable effects in the neutral iron signal, which are amplified by its possible condensation on the nightside. 
Here, we study the presence of molecular signals during a transit of \wasp\ observed with the CARMENES spectrograph and compare the contributions from this planet's morning and evening terminators. The results are somewhat puzzling, with formal detections of water vapor (5.5$\sigma$) and hydrogen cyanide (5.2$\sigma$) but at significantly different positions in the \kp-$\varv_{\textrm{sys}}$ diagram, with a blueshift of -14.3\,$\pm$\,2.6\,\kms\ and a redshift of $+$20.8\,$^{+7.8}_{-3.9}$\,\kms\ respectively, and a higher \kp\ than expected. The \h2o\ signal also appears stronger later on in the transit, in contrast to that of HCN, which seems stronger early on. We tentatively explain this by silicate clouds forming and raining out on the nightside of the planet, partially removing oxygen from the upper atmosphere. For atmospheric C/O values between 0.7 and 1, this leads to the formation of HCN at the planet's morning limb. At the evening terminator, with the sequestered oxygen being returned to the gas phase due to evaporation, these C/O values lead to formation of H$_2$O instead of HCN. Overall, if confirmed, these observations indicate that individual molecules trace different parts of the planet atmosphere, as well as nightside condensation, allowing spatial characterization. As these results are based on a single transit observation, we advocate that more data are needed to confirm these results and further explore these scenarios.}
\keywords{planets and satellites: atmospheres -- planets and satellites: gaseous planets -- planets and satellites: individual: \wasp}
\maketitle
%

\section{Introduction}
\label{Intro}

Ultra-hot Jupiters (UHJs) closely orbit their host starts in tidally locked configurations, with the
permanent stellar irradiation elevating their dayside temperatures well above 2000\,K  \citep{tan2019atmospheric}. 
Under such conditions, most of the molecular species are dissociated and atomic compounds become ionized \citep{hoeijmakers2019spectral, yan2019ionized, casasayas2019atmospheric, stangret2020detection, ehrenreich2020nightside, tabernero2021espresso}. This yields large contrasts in temperature, mean molecular weight, and atmospheric scale height between the day and the nightside of these planets \citep{komacek2016atmospheric, bell2018increased, arcangeli2018h, parmentier2018thermal, lothringer2018extremely, kitzmann2018peculiar, wardenier2021decomposing, savel2021no}.

A remarkable feature for UHJs is the large change in the viewing angle from the Earth during their transits. For instance, in the case of \wasp, this change is $\sim$30\textdegree. Because of this, the atmospheric regions probed through the leading and trailing limbs change significantly during the observations.
 \citet{ehrenreich2020nightside}, studying neutral iron absorption signals, recently showed that this effect impacts the appearance of compounds in \wasp.
High-dispersion Doppler spectroscopy was used to reveal the time-varying signal, as the radial velocity of the planet changes with respect to the Earth  during the transit. The iron absorption signal was found to be significantly greater from the evening side of the planet than from the morning side. 
Such time-resolved absorption signals also allow the atmospheric dynamics to be spatially resolved, with their net Doppler shift being a combination of the planet's rotation and high-altitude winds. During the first part of the transit, the combination of the rotational velocity (about $+$5\,\kms, redshifted) and the mean wind velocity at the layers probed with iron in the morning terminator (about $-$5\,\kms, blueshifted) results in a null Doppler shift of the signal in the rest-frame of the planet. As the transit progresses, the dayside of the evening terminator becomes observable, with the dayside of the morning terminator rotating out of view, causing the iron signature to be progressively blueshifted as the rotational and wind velocities add up to about $-$10\,\kms\ in the former. Here, we refer to the asymmetries observed in absorption signals in UHJs as the ``Ehrenreich effect'', regardless of their origin.

Interestingly, the zero-velocity component of the iron absorption disappears during the second half of the transit, when only the nightside of the morning terminator is probed. This was interpreted by \citet{ehrenreich2020nightside} as being caused by condensation in the nightside of \wasp.
However, scale-height differences between the morning and the evening limbs have been shown to reproduce the asymmetric iron signal without necessarily invoking iron condensation \citep{wardenier2021decomposing}. Moreover, \citet{savel2021no} claim that optically thick clouds are likely to form in the morning limb, effectively increasing the blueshift of the iron signal over transit.

Here, we aim to study the time dependence of molecular signals in order to further assess the spatial characterization of \wasp\ using high-dispersion cross-correlation spectroscopy \citep{snellen2010orbital, brogi2012signature, brogi2016rotation, nugroho2017high, nugroho2021first, alonso2019multiple, sanchez2019water, sanchez2020discriminating, guilluy2019exoplanet, hoeijmakers2020high, kesseli2020search, giacobbe2021}.
In Sect.\,\ref{obs_data}, we describe the observations and the methods we used to perform data reduction, telluric and stellar corrections, cross-correlation, and assessment of the significance of signals. In Sect.\,\ref{results} we present the main results of this work and their interpretation and, finally, in Sect.\,\ref{conclusions}, we discuss the main conclusions of this work.

\section{Observations and data analysis} \label{obs_data}

\begin{figure}
\includegraphics[angle=0, width=1\columnwidth]{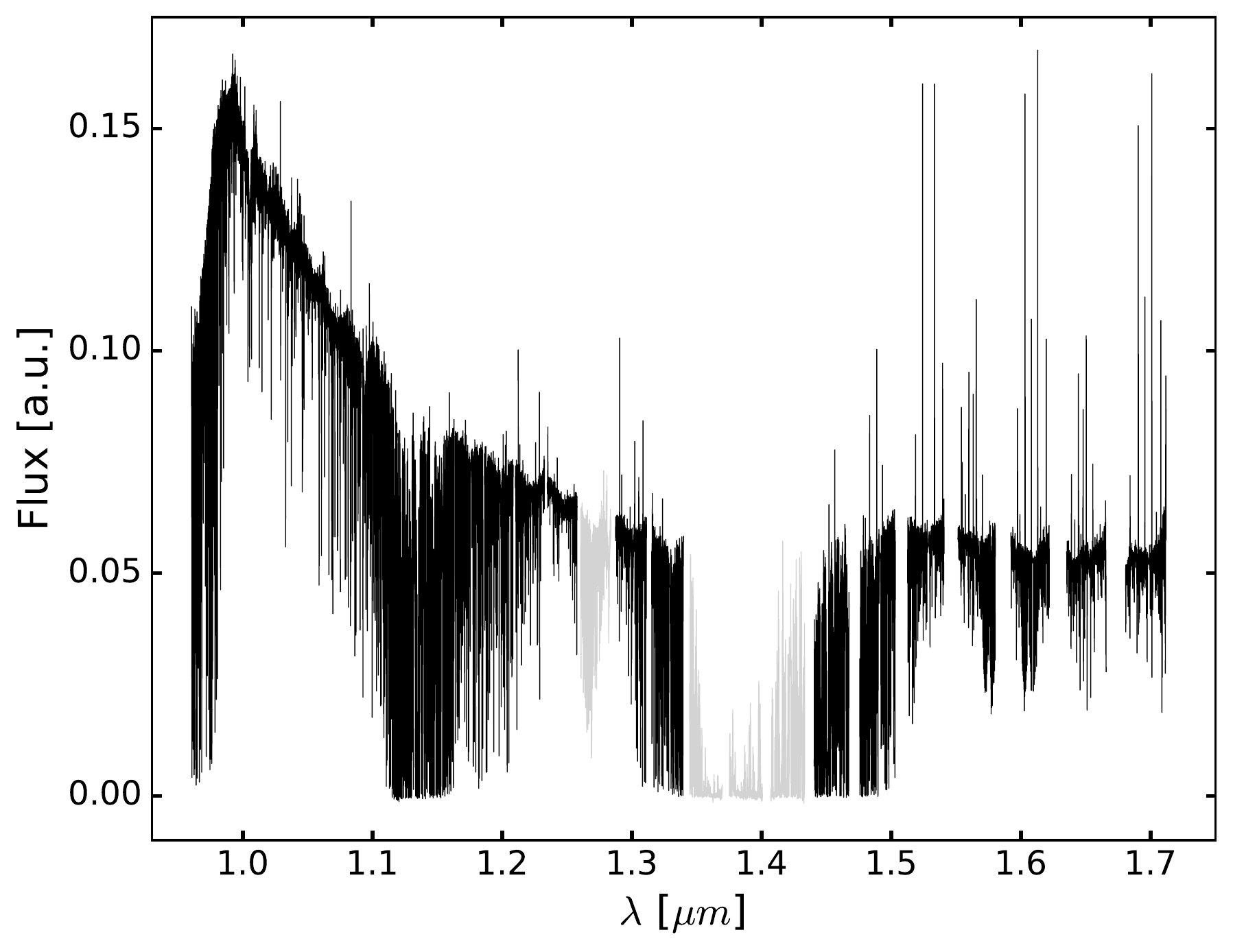} 
\caption{Stellar spectrum observed at an orbital phase of 0.007 (about mid-transit). Orders in black were included in the analysis. Orders in light gray were discarded because there is almost no flux around $\sim$\,1.40\,$\mu$m and because of the residuals in the final results around $\sim$\,1.27\,$\mu$m.}  
\label{fig:orders}
\end{figure}

We used publicly available CARMENES\footnote{Calar Alto high-Resolution search for M dwarfs with exoEarths with NIR and optical Échelle Spectrographs at the 3.5\,m Calar Alto Telescope.}\citep{quirrenbach2016carmenes, quirrenbach2018carmenes} data of \wasp\ from the Calar Alto Archive originally from proposals with CAHA internal programme number 022. The data were already reduced and pre-processed by the {\tt caracal} pipeline \citep{zechmeister2014flat, caballero2016carmenes}, which performs a bias and flat field correction and a wavelength calibration to provide spectra in the Earth's rest-frame with wavelengths in vacuum. We used the data from the CARMENES near-infrared (NIR) channel (0.96--1.71\,$\mu$m interval) at a spectral resolution of $\mathcal{R}$\,$\sim$\,80,400. The observations correspond to a transit of this UHJ that occurred on October 4, 2018, and is covered by 44 exposures of 498\,s with a mean signal-to-noise ratio (S/N) per pixel of $\sim$\,56. This same dataset was also recently used in \citet{landman2021} and \citet{casasayas2021carmenes} to perform atmospheric characterization of this UHJ. A second transit was observed on October 30, 2019, although with significantly poorer conditions (S/N\,$\sim$\,37),
which did not allow us to recover any planet signal (not shown). 

\begin{figure*}[htb!]
\centering
\includegraphics[angle=0, width=1.9\columnwidth]{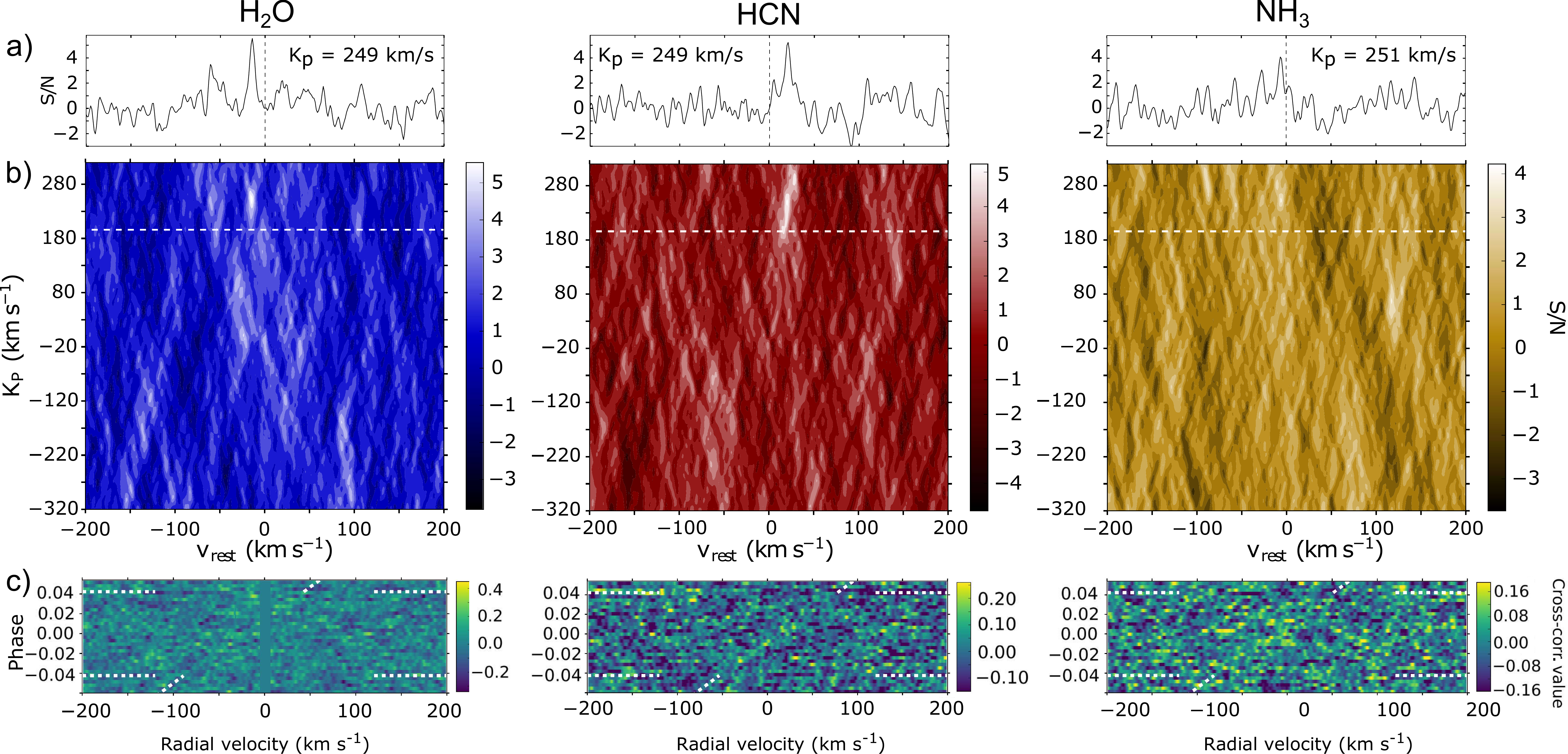}
\caption{Results of the cross-correlation analyses. (a) CCFs at the \kp\ presenting the maximum significance peak. (b) Cross-correlation maps (expressed as S/N) of potential signals for \h2o\ (left), HCN (middle), and \nh3\ (right) with respect to the exoplanet rest-frame velocity (horizontal axis) and \kp\ (vertical axis). The horizontal dashed lines mark the expected \kp\ of \wasp\ ($\sim$196\,\kms). (c) Cross-correlation matrices in the Earth rest-frame. The transit occurs between the horizontal dashed lines and the tilted dashed lines mark the velocities of the highest significance signal. Larger versions of the cross-correlation matrices in the Earth's rest frame and in the rest-frame of the exoplanet are shown in Appendix\,\ref{appendix:b}.} 
\label{fig:results}
\end{figure*}

We normalized the 28 NIR spectral orders using second-order
polynomial fits to their continuums, then corrected sky emission lines 5\% above the normalized continuum, and masked the intervals with the strongest telluric absorption (i.e., over 80\% of flux absorbed) by following the methods discussed in \citet{alonso2019multiple} and \citet{sanchez2019water, sanchez2020discriminating}. We excluded from all analyses the three spectral orders covering the center of the strongest water-vapor telluric band (1.34--1.43\,$\mu$m) where the flux received is too small (Fig.\,\ref{fig:orders}). Also, we found order 15 (1.233--1.283\,$\mu$m), which presents a strong telluric O$_2$ band, to introduce significant residuals in the final cross-correlation results and hence it was not used. In addition, we masked the $\pm$5\,\kms\ region in the cross-correlation maps of the water-vapor study to avoid the strongest telluric \h2o contamination near the Earth's rest frame (i.e., around 0\,\kms). 

The quasi-static telluric and stellar contributions dominating the spectra at this stage were corrected by applying the detrending algorithm {\tt SYSREM} \citep{tamuz2005correcting, mazeh2007transiting} order by order. The number of iterations required to remove these contributions and reach the level of planetary molecular absorption might vary between different spectral regions as the strength of telluric and stellar signals is wavelength dependent. In order to determine the necessary iterations for each order, we injected a model of the atmospheric absorption of the planet computed with {\tt petitRADTRANS} \citep{molliere2019petitradtrans} at the very beginning of the analysis and at the expected \kp--$\varv_{\textrm{sys}}$ for \wasp\ \citep[i.e., \kp \,=\,196.52\,$\pm$\,0.94\,\kms, $\varv_{\textrm{sys}}$\,$\sim$\,-1.167\,\kms;][]{ehrenreich2020nightside}. We assumed a pressure-temperature profile representative of the terminator region following the results of \citet{kataria2016atmospheric} for WASP-19\,b (see their Fig.\,3), which has a similar equilibrium temperature as \wasp. We investigated the presence of water vapor, hydrogen cyanide, ammonia, carbon monoxide, carbon dioxide, and methane in the atmosphere of \wasp, assuming constant volume mixing ratios (VMR; denoted as X) with pressure for simplicity. The line lists used were the default ones in {\tt petitRADTRANS}. For water vapor, we used $\log({\rm X_{H_2O}})$\,=\,$-$2.85, which was retrieved from \citet{edwards2020ares}. For the remaining undetected species in this planet, we assumed VMRs of $\log({\rm X_{HCN}})$\,=\,$-$5 (see Sect.\,\ref{results}), $\log({\rm X_{NH_3}})$\,=\,$-$6, $\log({\rm X_{CO}})$\,=\,$-$2, and $\log({\rm X_{CO_2}})$\,=\,$-$6, which are similar to the predicted abundances in the chemical equilibrium code of \citet{mollierevanboekel2016}.

In the \h2o analysis, the signal was injected at approximately three times the strength of the nominal model so that it was clearly recovered above the noise (see Fig.\,\ref{fig:injection_tests}) and, for each order separately, we determined the number of {\tt SYSREM} iterations that allowed us to maximize this retrieved injected signal in it \citep{birkby2017discovery, nugroho2017high, sanchez2019water, sanchez2020discriminating}. Consecutively, we ran the determined number of iterations for each order in the original spectra without injection. To avoid spurious results, we injected the signal in different \kp--$\varv_{\textrm{sys}}$ combinations and found similar results for all of them. 
Marginally worse results than those discussed in Sect.\,\ref{results} (within 1$\sigma$) were obtained by fixing 12 {\tt SYSREM} iterations for all spectral orders, avoiding the use of injections. The lower significance is expected given the very strong spectral dependence of the telluric \h2o contribution, which ensures that the analysis benefits from order-by-order optimization of the performance of {\tt SYSREM}.

For the other species, the weaker intrinsic strength of the absorption lines in most spectral orders and the rather low VMRs used prevented us from clearly recovering injections in the majority of them. Hence, we adopted a conservative approach in their analysis and used a fixed number of iterations for all spectral orders \citep{alonso2019multiple, stangret2020detection, nugroho2021first, cont2021detection}. We found the cross-correlation signals of HCN and \nh3\ to be maximized when applying seven and six {\tt SYSREM} iterations, respectively (see Sect.\,\ref{results}). No signal was observed with any number of iterations for CH$_4$, CO, and CO$_2$.

After the telluric correction, we cross-correlated the residual matrices with models of the expected atmospheric absorption of \wasp\ computed with {\tt petitRADTRANS} using the parameters discussed above in a range from $-$200 to 200\,\kms\ in steps of 1.3\,\kms\ and combined the cross-correlation results of all spectral orders separately for each molecule in order to enhance the detectability of possible signals (see Fig.\,\ref{fig:results}c). We then co-added in time the cross-correlation matrices for a range of orbital velocity semi-amplitudes (\kp) ranging from $-$280 to 280\,\kms. This allowed us to explore the presence of residuals at \kp\,$\sim$\,0\,\kms\ or spurious signals in the negative \kp\ space. At this stage, if a planet signal is detectable, it should reveal itself as a cross-correlation function (CCF) peak in the rest-frame of the  planet (see Fig.\,\ref{fig:results}b). In order to evaluate the significance of any possible peaks, we calculated the S/N of the CCFs at each \kp\ dividing the maximum value by the standard deviation in the rest of the velocity interval, excluding $\pm$15\,\kms\ around it.

\begin{figure*}[htb!]
\centering
\includegraphics[angle=0, width=1.9\columnwidth]{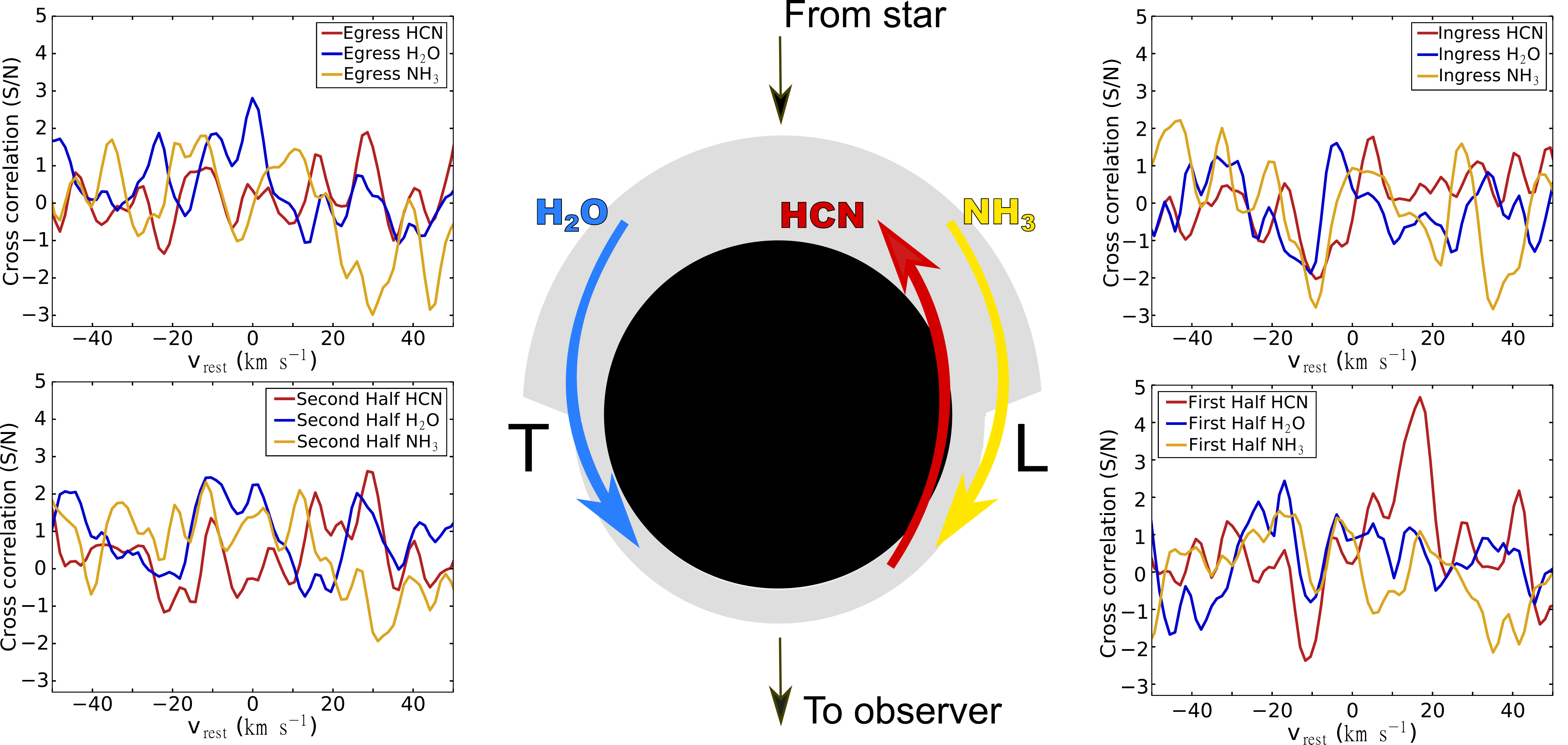}
\caption{Sketch of the asymmetric \wasp\ atmosphere and the spatial distribution of its constituents as suggested by the cross-correlation functions obtained for \h2o\ (blue), HCN (red), and \nh3\ (yellow) during ingress (top right), the first and second halves of the transit (bottom right and bottom left, respectively), and egress (top left). All CCFs were obtained at the expected \kp\ for \wasp\ (196\,\kms). The Doppler shift of the signals is indicated by the colored arrows in the sketch. That is, blueshifted signals (\h2o and \nh3) are marked by arrows pointing at the observer, whereas the redshifted HCN signal point towards the star. The leading and trailing limbs during the planet transit are indicated by the letters ``L'' and ``T'', respectively.}
\label{fig:sketch}
\end{figure*}

\section{Results and Discussion}
\label{results}

The resulting S/N maps (Fig.\,\ref{fig:results} and Table\,\ref{table:global_results}) show formal detections of water vapor (S/N\,$\sim$\,5.5) and hydrogen cyanide (S/N\,$\sim$\,5.2) and a tentative signal of ammonia (S/N\,$\sim$\,4.2). In contrast, we did not find any signals for CO, CH$_4$, or CO$_2$ (not shown). In particular, the S/N of the observed \h2o signal is consistent with injection recovery tests of models created using the retrieved abundance from Hubble Space Telescope observations \citep{tsiaras2018population,edwards2020ares}. For \h2o and \nh3, the \kp\ uncertainties do not encompass the expected velocity, which is in line with the OH results of \citet{landman2021} and the tentative H I signal in \citet{kesseli2021b}. In the case of HCN, while the peak \kp\ is also located at $\sim$250\,\kms, the expected value is inside the large error bars produced by its multiple-peak configuration in the \kp--$\varv_{\textrm{sys}}$ map. Recently, \citet{wardenier2021decomposing} studied deviations from the expected \kp-$\varv_{\textrm{sys}}$ that can be produced by the interplay between planet rotation, extreme and spatially dependent dynamics, and condensation in UHJs. These can result in multiple CCF peaks at lower-than-expected \kp\ values. However, no combination of drag strength and presence (or absence) of metal oxides in their models  can produce the larger \kp\ of these molecular signals. If confirmed, this would indicate that
additional sources of dynamics or stronger chemical or temperature contrasts  possibly have to be included in future studies using global circulation models (GCMs).

\begin{table}[b!]
\centering
{\tiny
\caption{\label{table:global_results} S/N, \kp\ values, and rest-frame velocities of the maximum significance signals.}
\begin{tabular}{ccccc} 
\hline
\hline
\noalign{\smallskip}
Molecule & S/N & \kp\ (\kms) & $\varv_{\rm rest}$ (\kms) & FWHM (\kms)$^a$\\
\noalign{\smallskip}
\hline
\noalign{\smallskip}
\h2o\ & 5.5 & 249\,$\pm38$ & --14.3\,$\pm2.6$ & 6.2\,$\pm1.3$\\
  \noalign{\smallskip}
HCN & 5.2 & 249\,$^{+51}_{-56}$ & $+$20.8\,$^{+7.8}_{-5.2}$ & 6.7\,$\pm1.5$\\
 \noalign{\smallskip}
\nh3\ & 4.2 & 251\,$\pm12$ & --6.5\,$\pm+2.6$ & 6.5\,$^{+1.8}_{-1.2}$\\ 
\noalign{\smallskip}
\hline
\end{tabular}
\tablefoot{Uncertainties correspond to $\pm1$ S/N. $^a$ FWHM measured at the peak \kp\ for \h2o and \nh3, for which there is no statistically significant signal at the expected \kp.}
}
\end{table}

We find the three signals to present similar full widths at half maximum (FWHM; see Table\,\ref{table:global_results}). We note that the reported FWHM for water vapor and ammonia were measured at the peak \kp, as no statistically significant signal was observed at the expected value for \wasp. However, for hydrogen cyanide,  the FWHM could be measured at the expected \kp\,$\sim$\,196\,\kms. All signals present consistent widths ranging from 6.2\,\kms\ to 6.7\,\kms. These broadenings are larger than expected from the instrumental function only because the resolution element is $\sim$\,3\kms, which points at strong dynamics in the atmosphere of \wasp. We note that these signals are significantly narrower than that of OH in \citet{landman2021} or those measured for atomic species in \citet{ehrenreich2020nightside}, \citet{tabernero2021espresso}, and \citet{kesseli2021b}. This is likely because these molecules are expected to be confined to lower atmospheric regions than those probed by atomic absorption and also OH \citep[see, e.g., Fig.\,1 in ][]{landman2021}. At these higher pressures, strong day-to-nightside or night-to-dayside winds \citep[see, e.g.,][]{wardenier2021decomposing} can shift the signals significantly without producing a large broadening. On the contrary, the higher atmospheric layers have been shown to present vertical winds or outflows \citep[see, e.g.,][]{seidel2021into, kesseli2021b} that move towards and also away from the observer, causing a small shift but a significant broadening of the spectral lines. However, it is important to caution that the measurement of the FWHM might be complicated by potential asymmetries of the planet trail in the cross-correlation matrix. After co-adding the CCFs in time, such asymmetries could translate into broad bumps or multiple peaks of lower significance in the 1D CCF. In such cases, the physical information contained in the retrieved FWHM might not be fully reliable.

The signals show significantly different Doppler shifts with respect to the planet rest-frame velocity (i.e., $\varv_{\rm rest}$\,=\,0\,\kms). The \h2o\ signal is blueshifted by -14.3\,$\pm$\,2.6\,\kms, indicative of a strong day to nightside wind at the atmospheric pressures probed by this molecule. This Doppler shift is in agreement with that observed for neutral iron towards the end the transit \citep{ehrenreich2020nightside} and for OH (Landman et al. 2021).
However, the HCN CCF peak shows a puzzling redshift of $+$20.8\,$^{+7.8}_{-5.2}$\,\kms\ at the peak \kp, with rather wide uncertainty intervals due to the large region of high significances in the map. Recently, a region of rapid atmospheric expansion in \wasp\ was proposed by \citet{seidel2021into}, who found vertical winds of 22.7\,$^{+{4.9}}_{-4.1}$\,\kms\ with an in-depth study of the broadening of Na lines \citep{seidel2019hot, tabernero2021espresso}. However, we do not observe such a strong broadening in the 1D CCF of HCN. If real, the redshifted signal is indicative of an atmospheric region moving away from the observer at lower atmospheric layers, without further indications of vertical winds.

According to atmospheric circulation studies of \wasp\ in \citet{wardenier2021decomposing} and \citet{savel2021no}, a strong atmospheric expansion away from the observer and winds flowing from the nightside to the dayside are both possible in the morning terminator of this planet, and the latter are the strongest in low- or no-drag scenarios. Thus, the HCN signal could be arising from a flow towards the dayside atmosphere at the morning terminator, where the irradiated atmosphere (i.e., larger scale height) expands away from the observer. This scenario would be of special importance because, among other reasons, it would provide evidence against the presence of thick clouds in the morning limb. Such thick clouds have been proposed as a possible explanation for the asymmetric iron absorption observed in \citet{ehrenreich2020nightside} in the simulations of \citet{savel2021no}. However, HCN should be confined to lower atmospheric layers than iron and, thus, confirmation of this signal would effectively rule out iron masking in this limb by clouds, favoring other origins for the asymmetric signals (e.g., strong temperature contrasts between limbs, iron condensation across the nightside, or a combination of both). Regarding the \nh3\ signal, its lower significance does not allow us to draw robust conclusions. Both the presence of \nh3\ and its blueshift need to be further studied with future observations. 

In order to explore the different atmospheric regions, we investigated the time-dependence of the signals. Figure\,\ref{fig:sketch} shows a basic sketch of the planet atmosphere and the CCF results we obtained during ingress, egress, and half transits assuming the expected \kp\ of \wasp. During ingress and egress, no significant signals (>$4\sigma$) are observed for any molecule. As both ingress and egress last for 0.01639\,d, about three spectra each in our case, the absence of signals in these intervals could
simply be explained by an overly low S/N.
By extending the analysis to the first half of the transit (orbital phases, $\phi$\,$<$\,0), with both the leading and the trailing limbs being fully observable in approximately the first 12 in-transit spectra, we observe HCN at a S/N$\sim$4.6.
No clear \h2o signal is observed in this time interval at the expected \kp\ (upper limit S/N\,$\sim$\,2.5).

\begin{figure}[t!]
\centering
\includegraphics[angle=0, width=0.8\columnwidth]{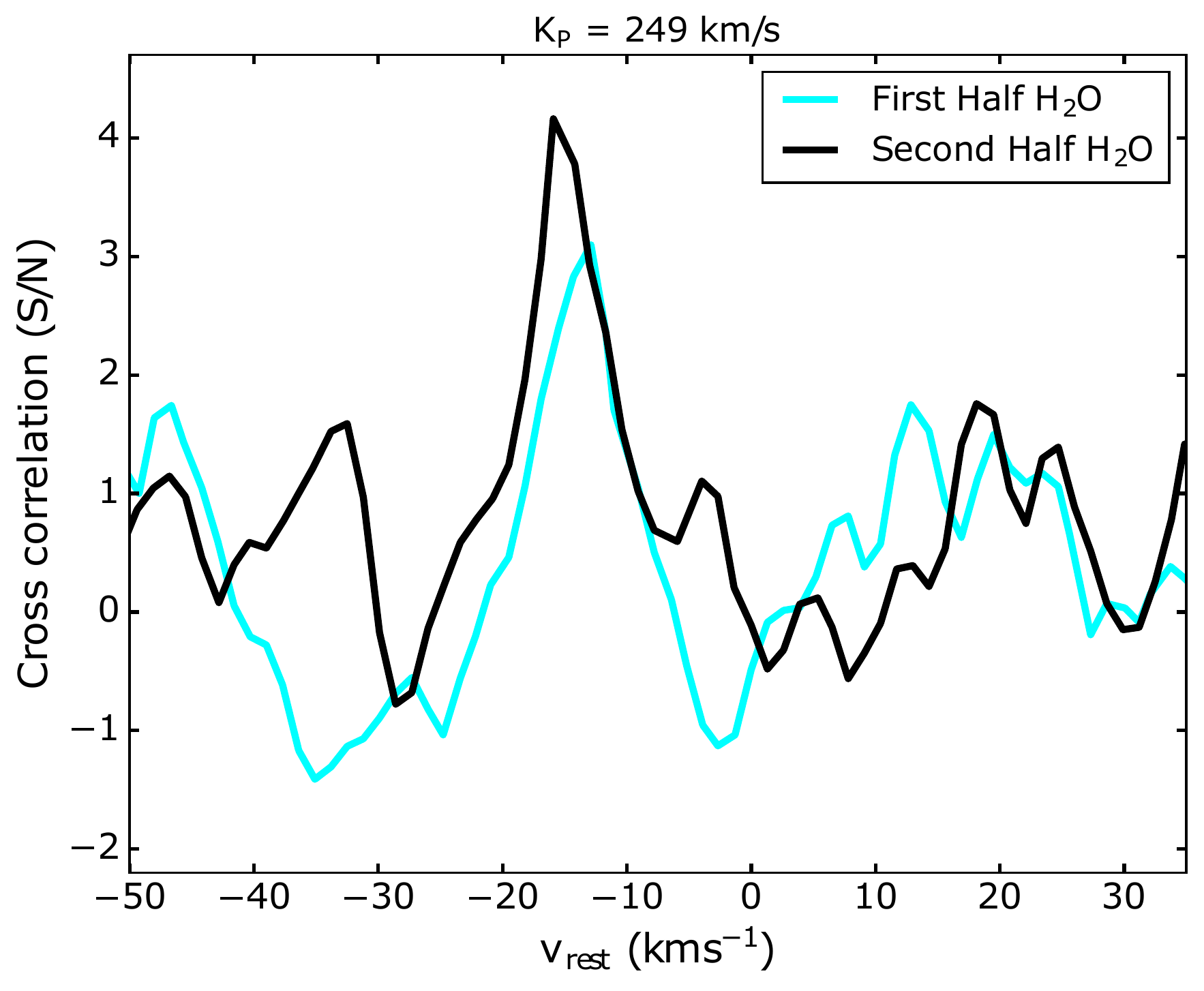}

\includegraphics[angle=0, width=0.8\columnwidth]{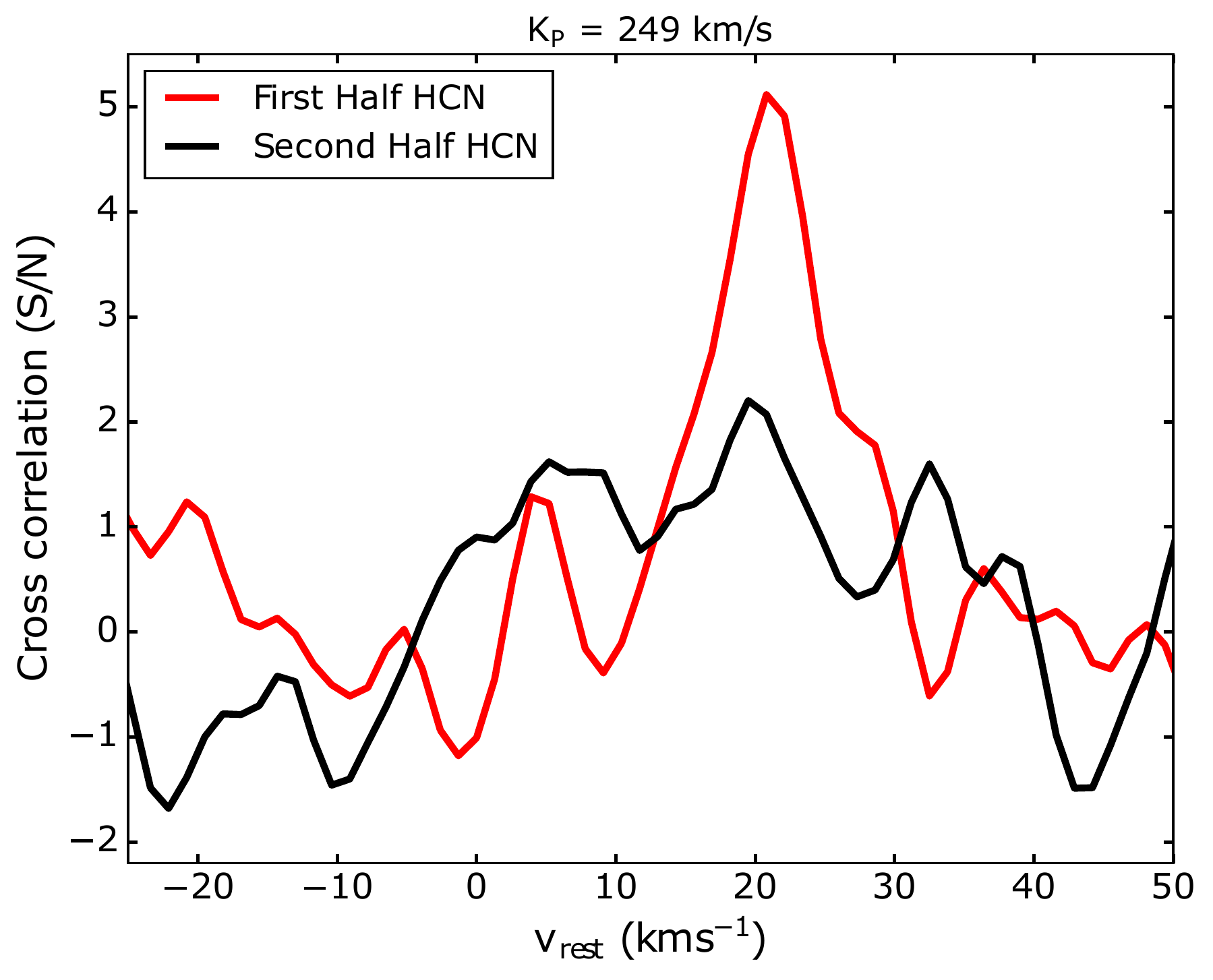}
\caption{Comparison of CCF signals in the first and second half of the transit for \h2o (top) and HCN (bottom) at the \kp\ of maximum significance for both molecules (249\,\kms).} 
\label{fig:halves_comp}
\end{figure}

During the second half of the transit ($\phi$\,$>$\,0), the HCN signal is no longer detected. Remarkably, no significant \h2o signal can be observed at the expected \kp\ in the second half of the transit either. In this case, we would expect CCF signals at the expected \kp\ to be weaker and broader than what is observed at the peak \kp. Potentially, there is a broad double-peaked CCF bump in the second half of the transit at roughly $\varv_{\rm rest}$ ranging from $-$15 to 5\,\kms, but it is not possible to confirm this because of the low S/N. By studying the two halves at the peak \kp\ (249\,\kms, see Fig.\,\ref{fig:halves_comp}), we observe a slightly stronger \h2o absorption in the second half of the transit. Although the observational evidence is low, this could be due to a larger contribution from the dayside part of the evening terminator at $\phi$\,$>$\,0. As for HCN, the same conclusions can be extracted from both Figs.\,\ref{fig:sketch} and \ref{fig:halves_comp}, with the signal being stronger in the first half. Interestingly, sodium and potassium, also predicted to be more abundant in the cooler nightside, are found to present stronger signals in the first half of the transit in \citet{kesseli2021b}. 

\begin{figure}[t!]
\centering
\includegraphics[angle=0, width=1.\columnwidth]{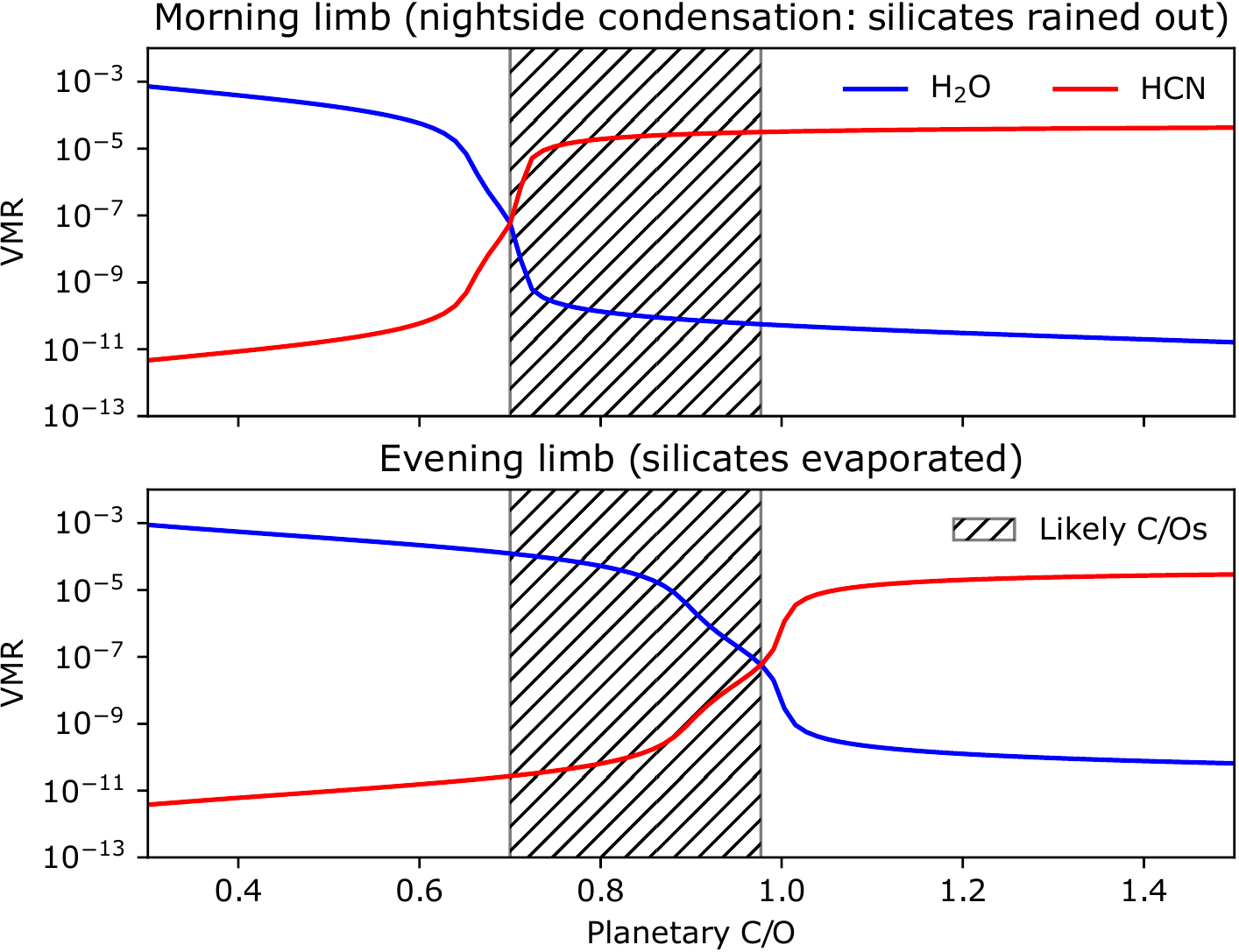}
\caption{Predicted VMRs of \h2o (blue solid lines) and HCN (red solid lines) in the morning ({\it upper panel}) and evening ({\it lower panel}) terminators of WASP-76b as a function of the total (gas+solid phase) planetary C/O. For the upper panel, the removal of oxygen into deep, rained-out silicate clouds has been taken into account. The hatched area indicates the C/O value range where \h2o and HCN can occur simultaneously (but at opposing limbs).} 
\label{fig:easyCHEMtest}
\end{figure}

From a thermochemical perspective, the concurrent detection of \h2o and HCN is unexpected: at temperatures $\gtrsim 1000$~K, \h2o occurs for C/O values $\lesssim 1$, whereas HCN requires C/O values $\gtrsim 1$ \citep[see Fig.\,\ref{fig:easyCHEMtest} and, e.g.,][]{loddersfegley2002,madhusudhan2012,mollierevanboekel2015}. The transition between both scenarios is quite sharp, such that this simultaneous detection requires significant fine tuning of the C/O ratio, which we deem unlikely. Instead we propose that silicates might condense and rain out on the planet's nightside, partially removing oxygen from the gas phase. For total (gas+solids) C/O values between 0.7 and 1, this effectively elevates the gas-phase C/O to values $\gtrsim 1$. 
Therefore, if the sequestered oxygen is not released immediately by silicate evaporation once the atmospheric flow crosses the morning terminator, this naturally leads to the formation of HCN. By the time the evening terminator is reached, the deep silicates have been evaporated on the hot dayside, and the increased oxygen abundance favors the formation of \h2o and the destruction of HCN. This mechanism would have the benefit of requiring much less fine tuning of the atmospheric C/O; any value between 0.7 and 1 suffices.

To demonstrate this chemical behavior, we carried out the following test: using the equilibrium chemistry code described in Appendix A.2 of \citet{mollierevanboekel2016}, we modeled both terminators probed by adopting a pressure of 0.001 bar, and the planet's equilibrium temperature of 2230~K. We then investigated the expected atmospheric \h2o and HCN VMRs at the evening terminator as a function of the total (gas+solids) atmospheric C/O ratio. For the morning terminator, we ran an additional calculation for every tested total C/O ratio. Namely, we set the atmospheric temperature and pressure to 1000 K and 100 bar in order to mimic the conditions at the cloud bases of the nightside of the planet. We note that 100~bar is much higher than the pressures probed during transits or the photospheric location of a planet, even for emission. However, this value was simply chosen for convenience, as it guarantees that all species of interest have condensed. At a temperature of $1000$~K, lower pressures (even below 1~mbar) would also have sufficed to guarantee silicate condensation. From the resulting low-temperature equilibrium abundances, we then calculated the amount of oxygen that was locked up in the condensates (among them MgSiO$_3$, Mg$_2$SiO$_4$, MgAl$_2$O$_4$, Al$_2$O$_3$, etc.). This amount was then subtracted from the oxygen available for gas phase chemistry at the morning terminator, calculated at 0.001 bar, and 2230~K. As can be seen in Figure \ref{fig:easyCHEMtest}, this treatment naturally results in a scenario where \h2o is expected only at the evening terminator, while HCN is expected only at the morning terminator, over a wide range of C/O values ($\sim$0.7 to $\sim$1). We note that condensation affecting the gas phase C/O and chemistry is a well-known effect \citep[e.g.,][]{fortney2006,hellingwoitke2014,mollierevanboekel2015}. Here we use it to explain the seemingly peculiar chemical behavior at the opposing limbs of the 
planet. 
If indeed the contributions from \h2o and HCN are confirmed to come only from the evening and morning limbs, respectively, then it is important to note that the planet's rotation will blueshift the water vapor signal by $-5$\,\kms\ and redshift the hydrogen cyanide signal by $+5$\,\kms. This would effectively reduce the actual wind speeds necessary to reach the measured Doppler shifts.

\section{Conclusions} 
\label{conclusions}

The different circumstances on the day and the nightsides of ultra-hot Jupiters produce significantly different atmospheric conditions in terms of scale height, composition, and overall dynamics. Such variability can translate into highly asymmetric absorption signals observed during primary transits (i.e., the Ehrenreich effect), as has recently been reported for \wasp\ in studies of neutral iron by \citet{ehrenreich2020nightside} and \citet{kesseli2021confirmation}. We report evidence of asymmetric molecular absorption features in a single transit observation of \wasp.
The main findings of this work can be summarized as follows:
\begin{itemize}
    \item We confirm the previously reported \h2o detection in \wasp\ \citep{tsiaras2018population, edwards2020ares} using high-resolution spectra. The strength of our signal is consistent with that of these latter studies. This CCF signal is blueshifted, suggesting strong winds flowing from the day to the nightside, which is in line with Fe and OH observations \citep{ehrenreich2020nightside, landman2021}. The \h2o signal is marginally stronger in the second half of the transit, which might point to a larger contribution from the evening terminator.
    \item We present evidence of nitrogen chemistry in \wasp. The redshifted HCN signal indicates that this molecule flows from the night to the dayside.
    Having detected \h2o and assuming chemical equilibrium, the HCN concentration should be much lower because \h2o and HCN are unlikely to co-exist. However, we propose that HCN forms close to the planet's morning limb. There, partial oxygen sequestration into rained-out clouds in the nightside would naturally explain its formation. This is supported by the HCN signal disappearing in the second half of the transit, when the morning terminator is no longer probed: as the condensates evaporate, \h2o is formed and HCN is destroyed. If confirmed, the HCN signal would provide evidence of a clear morning terminator in \wasp.
    \item In the three cases (i.e., \h2o, HCN, and \nh3), the maximum significance CCF peaks are consistently displaced towards higher-than-expected \kp\ values, which is in line with results presented in \citet{landman2021} and \citet{kesseli2021b}. The origin of the asymmetries producing the Ehrenreich effect in these signals is still uncertain as the latest GCM calculations for this planet do not predict such scenarios. Future observations at higher S/N should be able to shed light on these discrepancies, providing more information for future GCM analyses aiming to disentangle the complex dynamics arising in the atmosphere of UHJs.
\end{itemize}

\begin{acknowledgements}
We thank Joost Wardenier for helpful discussions. We also thank the anonymous referee for their insightful reading of the manuscript and very useful suggestions. We acknowledge funding from the European Research Council under the European Union's Horizon 2020 research and innovation program under grant agreement No 694513. P.M. and acknowledges support from the European Research Council under the European Union’s Horizon 2020 research and innovation program under grant agreement No. 832428-Origins. This research has made use of the Spanish Virtual Observatory (http://svo.cab.inta-csic.es) supported by the MINECO/FEDER through grant AyA2017-84089.7.
\end{acknowledgements}

\bibliographystyle{aa} 
\bibliography{references}
\appendix
\section{Injection recovery order by order for \h2o}
\label{appendix:a}
\begin{figure*}[ht!]
\centering
\includegraphics[angle=0, width=0.6\columnwidth]{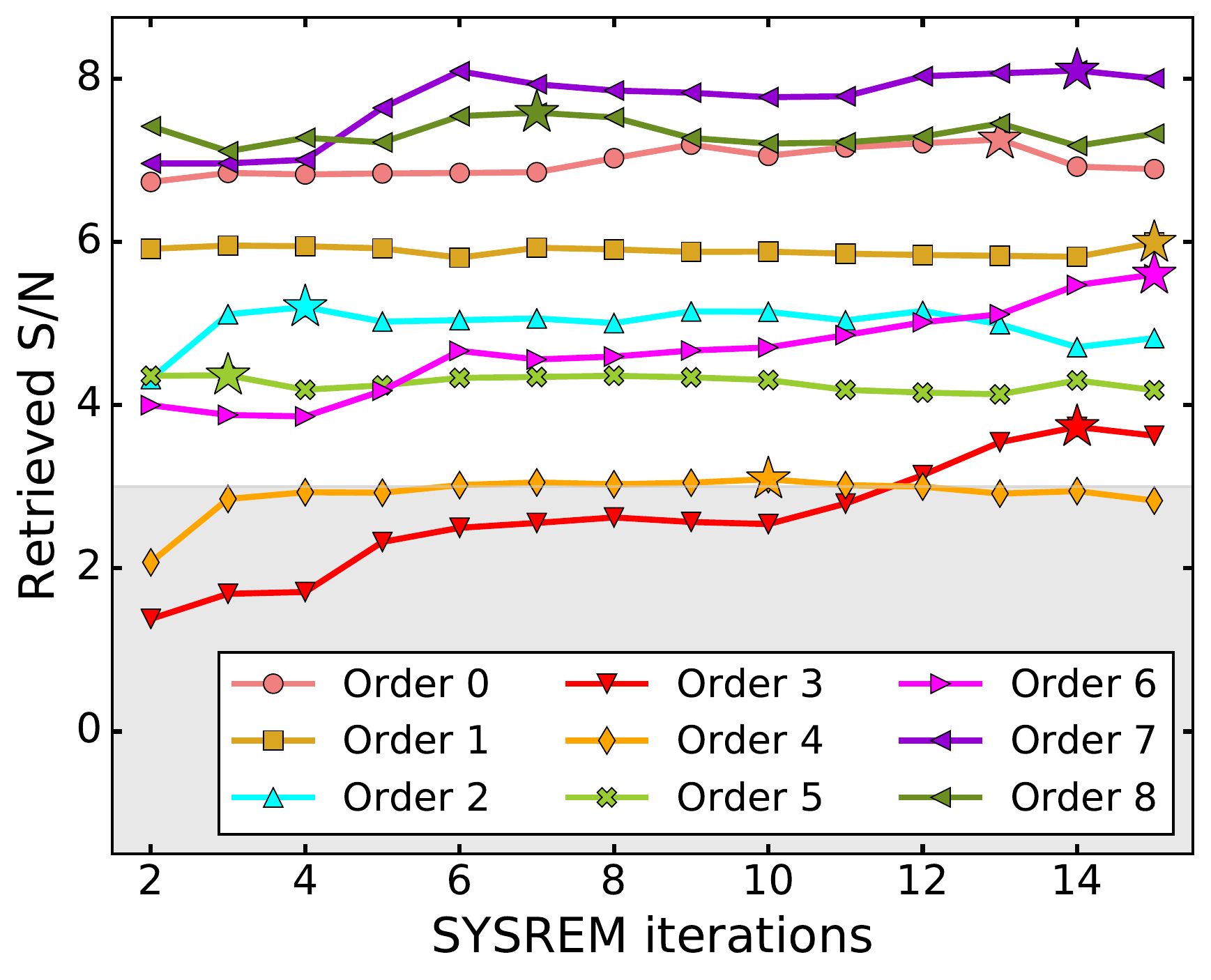}\includegraphics[angle=0, width=0.6\columnwidth]{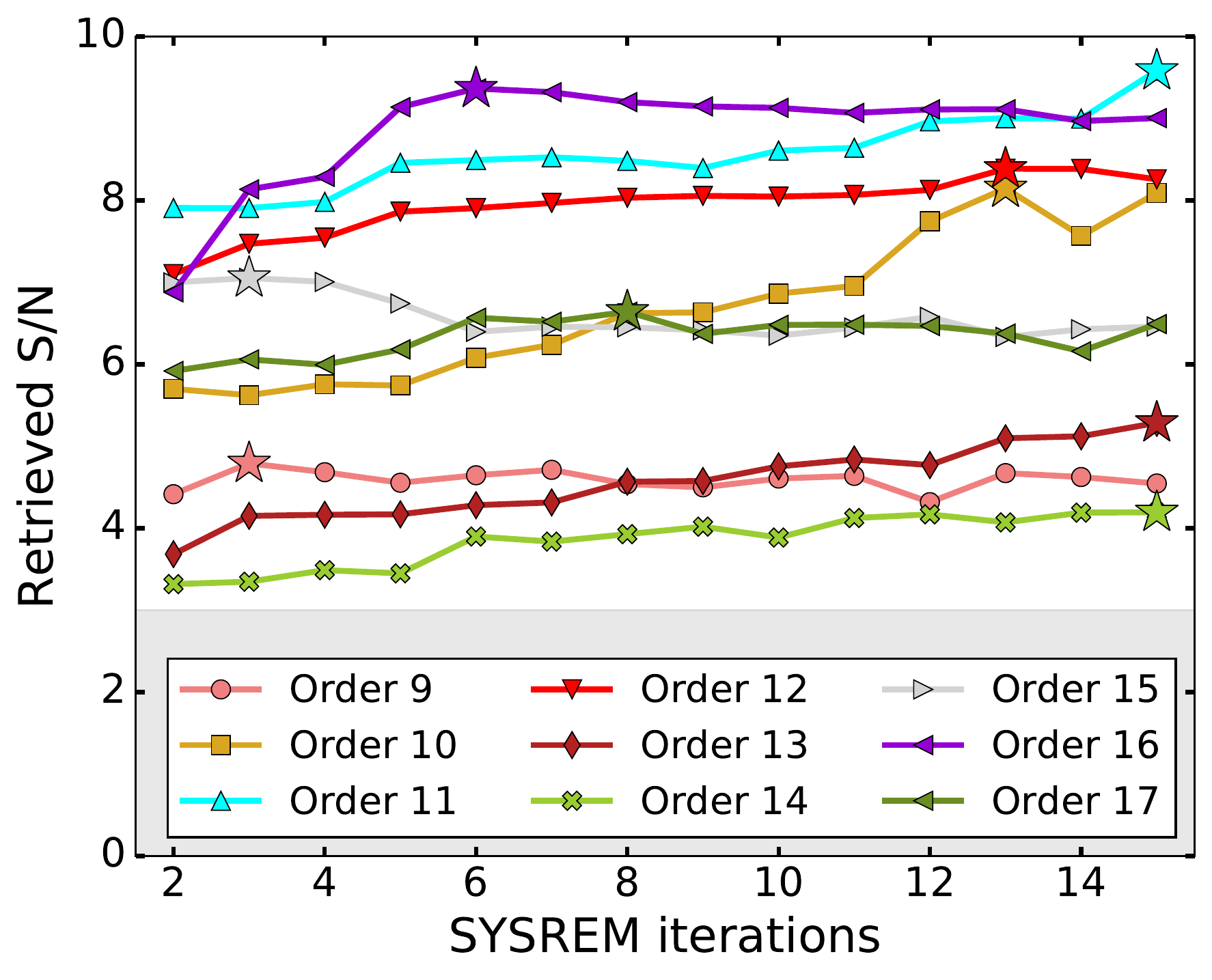}\includegraphics[angle=0, width=0.6\columnwidth]{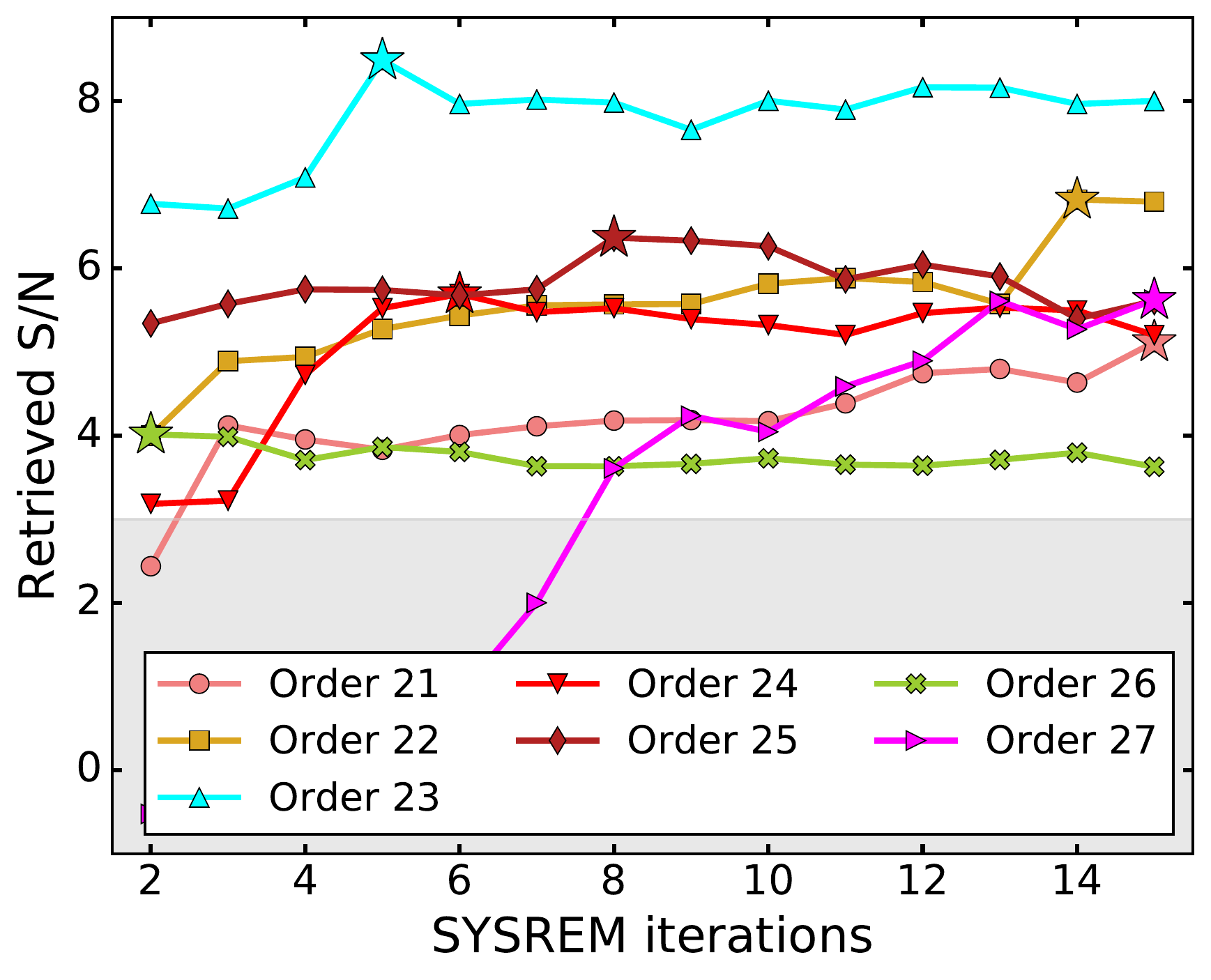}
\caption{Evolution of the S/N of the retrieved \h2o\ injected signal with increasing {\tt SYSREM} iterations. The model was injected at three times the expected level of absorption. The star symbols mark the iteration in which the recovery of the injected signal is maximized. The injected signal would not be recovered with confidence if its peak S/N was below 3 (shaded region). Spectral order 15 (in gray) was excluded from the analysis due to the presence of residuals in the final cross-correlation results.} 
\label{fig:injection_tests}
\end{figure*}

\section{Cross-correlation matrices}
\label{appendix:b}
\begin{figure*}[ht!]
\centering
\includegraphics[angle=0, width=1.3\columnwidth]{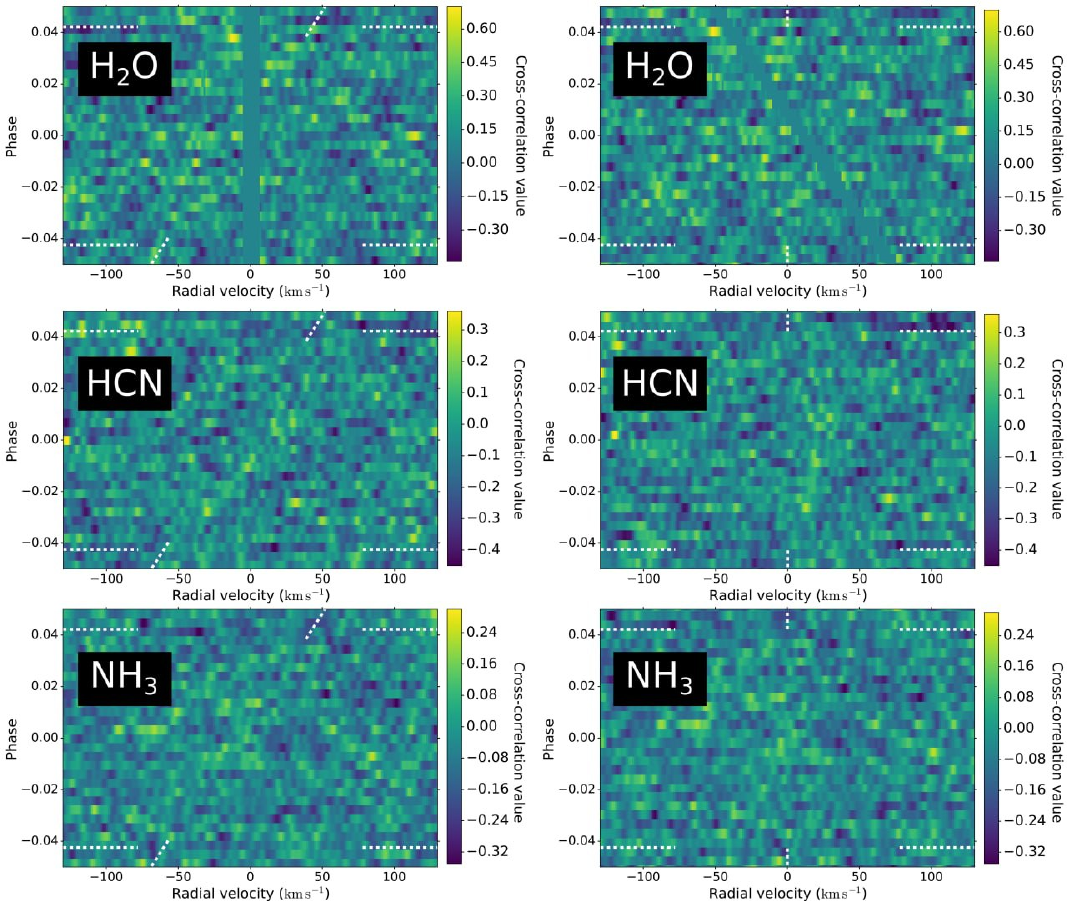}
\caption{Cross-correlation matrices obtained by following the methods discussed in Sect.\ref{obs_data}. Left column: Cross-correlation matrices in the Earth rest frame. The tilted dashed lines mark the exoplanet velocities. Right column: Cross-correlation matrices in the exoplanet rest-frame (\kp $\sim$\,196\,\kms). In this case, the exoplanet velocities are constantly 0\,\kms (vertical dashed lines). In all plots, the transit occurs between the horizontal dashed lines.} 
\label{fig:cc_matrices}
\end{figure*}

\end{document}